\documentclass[aps,prb,twocolumn,superscriptaddress,nofootinbib,amssymb]{revtex4-1}
\usepackage{graphicx}
\usepackage{amsmath}
\usepackage{times}
\usepackage{tikz}
\usetikzlibrary{arrows,automata}
\usepackage{multirow}
\usepackage{float}
\usepackage{subfig}
\usepackage{tabularx}
\graphicspath{{main_figures},{suppl_figures},{Additional_figures}}
\usepackage[colorinlistoftodos]{todonotes}
\usepackage{comment}
\usepackage{caption}
\begin{document}

\title{Staggered-flux state for rectangular-lattice spin 1/2 Heisenberg antiferromagnets}

\author{N. E. Shaik}\affiliation{Ecole Polytechnique F\'ed\'erale de Lausanne (EPFL), Institute of condensed matter physics, CH-1015 Lausanne, Switzerland}

\author{B. Dalla Piazza}\affiliation{Ecole Polytechnique F\'ed\'erale de Lausanne (EPFL), Institute of condensed matter physics, CH-1015 Lausanne, Switzerland}

\author{D. A. Ivanov}\affiliation{Institute for Theoretical Physics, ETH Z\"urich, CH-8093 Z\"urich, Switzerland }

\author{H. M. R{\o}nnow}\affiliation{Ecole Polytechnique F\'ed\'erale de Lausanne (EPFL), Institute of condensed matter physics, CH-1015 Lausanne, Switzerland}

\date{\today}

\begin{abstract}
We investigate the spin-1/2 Heisenberg  model on a rectangular lattice, using the Gutzwiller projected variational wave function known as the staggered flux state. Using Monte Carlo techniques, the variational parameters and static spin-structure factor for different coupling anisotropies $\gamma=J_y/J_x$ are calculated. We observe a gradual evolution of the ground state energy towards a value which is very close to the 1D estimate provided by the Bethe ansatz and a good agreement between the finite size scaling of the energies. The spin-spin correlation functions exhibit a power-law decay with varying exponents for different anisotropies. Though the lack of N\'eel order makes the staggered flux state energetically unfavorable in the symmetric case $\gamma=1$, it appears to capture the essence of the system close to 1D. Hence we believe that the staggered flux state provides an interesting starting point to explore the crossover from quantum disordered chains to the N\'eel ordered 2D square lattices.
\end{abstract}

\maketitle

Heisenberg interaction constitutes a major type of the magnetic interaction between the spins in many materials. It emerges naturally in highly correlated materials with a large Coulomb repulsion. In the field of quantum magnetism, models in various dimensions, involving both ferromagnetic and anti-ferromagnetic interactions, have extensively been studied. Perhaps one of the most interesting models is the two dimensional quantum square lattice Heisenberg Antiferromagnet which, despite its simplicity, lacks an exact analytical solution. One of the key motivations behind studying this model is  because it describes the functional building blocks  of parent compounds of high-temperature superconductors like cuprates\cite{Bednorz1986,Buchanan2001}, and magnetic fluctuations are speculated to be reason for the pairing mechanism of the cooper pairs\cite{Monthoux}. At zero temperature, the ground state has anti-ferromagnetic long-range order with algebraically decaying transverse correlations. The low energy excitation spectrum consists of magnon excitations described by spin wave theory\cite{Anderson2,Kubo}. However inelastic-neutron-scattering studies on these structures have shown a striking anomaly at the $(\pi,0)$ k-point that could not be explained using conventional spin wave theory\cite{Ronnow,Christensen}. The main feature of this anomaly is the loss of almost half of spectral weight in the magnon branch, which emerges as a high energy continuum.  

This issue led to the work by Dalla Piazza et al.\cite{DallaPiazza}, where  it was shown that the Staggered Flux (SF) state\cite{Marston}, a variant of RVB, is capable of capturing the essential features of the quantum anomaly. The interesting aspect of this work is that the observed excitations at the quantum anomaly are 2D analogues of particles carrying fractional ($S=1/2$) quantum numbers termed as `spinons'. Spinons are the fundamental excitations found in the exact 1D solution through Bethe ansatz\cite{Bethe} which have experimentally been observed\cite{Tennant,Lake2005,Mourigal}. In higher dimensions, they are proposed to be found mostly in frustrated lattices capable of hosting a quantum spin liquid\cite{Balents}, a disordered phase with high degeneracy at $T=0$.  On the other hand quasi 1D systems called spin ladders have exhibited experimental features where the low energy bands are magnon like and high energy features are spinon like\cite{Lake}. Contrary to the `spinon' picture, alternate proposal by Powalski et al\cite{Powalski,Powalski2018}, based on continuous similarity transformation of the Hamiltonian in momentum space,  attribute the $(\pi,0)$ anomaly to higher order magnon-magnon interaction denoted as magnon-Higgs-like scattering. 

 Motivated by the work by Dalla Piazza et al \cite{DallaPiazza}, we extend their methodology to the rectangular lattice, where by tuning the ratio of spin couplings in the two lattice directions we can interpolate between the square-lattice limit (where spinons are only conjectured at one wave vector in the magnon band) and one-dimensional chains (where spinons are elementary excitations at all wave vectors). In this work, preliminary studies of the projected staggered-flux wave functions on such rectangular lattices are reported. We compute the variational energies and the spin correlations as a function of the anisotropy parameter. We also comment on the finite-size effects which are especially important in the strongly anisotropic case. These results will be of future use for the analysis of spin excitations in such states.

 Our work is related to that of T.Miyazaki et al.\cite{Miyazaki}, who studied a similar variational ansatz in the Schwinger-boson construction. Due to the difference in the particle statistics (fermion vs. bosons) we do not expect exact agreement between the wave functions in their work and in ours. Furthermore, in view of applying our wave function to the spinon deconfinement problem along the lines of DallaPiazza et al.\cite{DallaPiazza}, we do not include antiferromagnetic ordering in our ansatz.

\section{Method}

\noindent 
We consider the Heisenberg Hamiltonian on the rectangular lattice
\begin{equation}\label{H-physical}
\mathcal{H} = \sum_{\left\langle i,j \right\rangle} J_{ij} \mathbf{S_i\cdot S_j}
\end{equation}
where $\mathbf{S_i,S_j}$ are the spin-1/2 operators on nearest-neighbor sites $\langle i,j\rangle$. The coupling $J_{ij}$ is equal to $J_x$ in the $x$ direction and to $J_y$ in the $y$ direction. Without loss of generality, we choose $J_y\leq J_x$ so that the anisotropy parameter $\gamma=J_y/J_x$ lies between 0 and 1.

Following the usual variational procedure for Gutzwiller-projected wave functions\cite{Gros,Dmitriev}, we consider the ground state $\left|\psi_{SF}\right\rangle$ of the auxiliary (``mean-field") Hamiltonian 
\begin{equation}
H_{SF}=-\sum_{\left\langle i,j \right\rangle, \sigma}
\chi_{ij} c^\dagger_{i\sigma}c_{j\sigma} \,
\label{H-mean-field}
\end{equation}
where $c^\dagger_{i\sigma}$ and $c_{i\sigma}$ are spin-1/2 fermion creation and annihilation
operators and the parameter $\chi_{ij}=J_{ij}\left\langle c^\dagger_{i\sigma} c_{j\sigma} \right\rangle$. We then optimize these parameters $\chi_{ij}$ within a certain symmetry class to minimize
the variational energy 
\begin{equation}
E=\left\langle GS \right| \mathcal{H} \left| GS \right\rangle
\end{equation}
of its projected ground state
\begin{equation}
\left| GS \right\rangle = P_{D=0} \left| \psi_{SF} \right\rangle \, ,
\label{projection}
\end{equation}
where the operator $P_{D=0}$ projects onto states with exactly one fermion per site.

We restrict our study to the staggered-flux ansatz for $\chi_{ij}$ (see, e.g., Refs\cite{Ivanov,marstonjb})
with different amplitudes in the $x$ and $y$ direction:
\begin{equation}
\chi_{i,i+x}=\chi_x e^{i(-1)^{i_x+i_y}\varphi/4}\, , \quad
\chi_{i,i+y}=\chi_y e^{-i(-1)^{i_x+i_y}\varphi/4}\, .
\label{staggered-flux}
\end{equation}

\begin{figure}[H]
    \centering
    \includegraphics[scale=0.5]{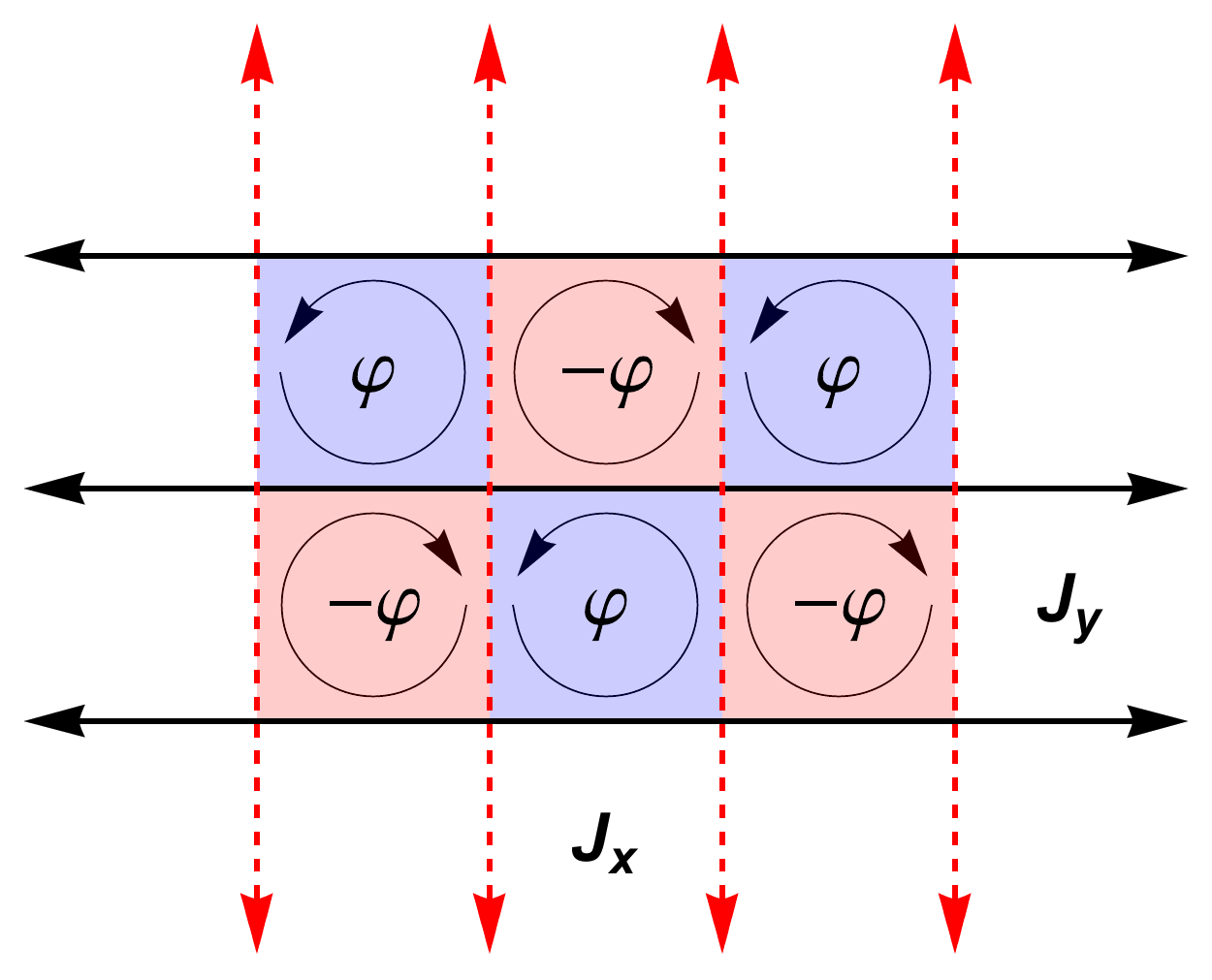}
    \caption{An illustration of staggered flux state with anisotropic couplings, represented as flux $\varphi$ threading the lattice in a staggered manner.}
    \label{stag_flux}
\end{figure}
Due to the projection, there is a redundancy in the phase of the fermion on
each site, and the projected state Eq.~(\ref{projection}) only depends on the total flux $\varphi$ through a lattice cell but not on the distribution of
the flux over phases of individual hopping amplitudes  $\chi_{ij}$ in Eq.~\eqref{staggered-flux}. Also, the overall normalization of $\chi_{ij}$ has no effect on the wave function. The variational wave function thus depends on two parameters: the flux
$\varphi$ and the hopping anisotropy
\begin{equation}
\alpha=\chi_y/\chi_x
\end{equation}

Note that there is a larger symmetry in the particle-hole space that makes
our projected staggered-flux state identical to a corresponding projected
d-wave superconducting state\cite{Bardeen}. For the same reason, the
projected staggered-flux state $\left| GS \right\rangle$ 
has a full translational symmetry, even though the unprojected state 
$\left| \psi_{SF} \right\rangle$ does not.

The spectrum of the auxiliary Hamiltonian Eq.~(\ref{H-mean-field}) is gapless
with nodes at the wave vector $(\pm \pi/2, \pm \pi/2)$ and energy given by:
\begin{equation}\label{Delta}
\varepsilon^\pm_k=\pm\frac{1}{2}\left| \chi_x e^{i\varphi/4}\cos k_x +\chi_y \ e^{-i\varphi/4}\cos k_y \right|\, .
\end{equation}
The ground state wave function corresponds to filling the lower band within the magnetic brillouin zone (MBZ) with up and down spins,
\begin{equation}
\left| \psi_{SF} \right\rangle = \prod_{k \epsilon MBZ} d_{k\uparrow -}^\dagger d_{k\downarrow -}^\dagger \left| 0 \right\rangle 
\end{equation}
where the operators $d$ ($d^\dagger$) are the annihilation (creation) operators for the eigenstates of the Hamiltonian Eq.~\eqref{H-mean-field}.

The observables in the projected state Eq.~(\ref{projection}) are calculated as
\begin{equation}\label{expectation_value}
\left\langle GS \right| O \left| GS \right\rangle = 
\frac{\left\langle \psi_{SF} \right| P_{D=0} O P_{D=0} 
\left| \psi_{SF} \right\rangle}{\left\langle \psi_{SF} \right|P_{D=0}\left| \psi_{SF} \right\rangle}
\end{equation}
Considering a Hilbert space made up of states $\left| \beta \right\rangle$  where all the lattice sites
are singly occupied, we may write the Gutzwiller projector as 
$P_{D=0}=\sum_{\beta} \left|\beta \right\rangle \left\langle \beta \right|$ and express the expectation value Eq.~\eqref{expectation_value} as a statistical average:
\begin{equation}\label{MCeq}
\left\langle  O   \right\rangle = \sum_\beta  \underbrace{\frac{\left|\left\langle \beta|\psi_{SF} \right\rangle \right|^2}{\sum_{\beta'} \left|\left\langle \beta'|\psi_{SF} \right\rangle \right|^2}}_{\rho(\beta)} \underbrace{\left (\sum_\gamma \left\langle \beta \right| O \left| \gamma \right\rangle \frac{\left\langle \gamma |\psi_{SF} \right\rangle}{\left\langle \beta|\psi_{SF} \right\rangle}\right)}_{f(\beta)}\, .
\end{equation}
The above equation has the form of a weighted average of a function $f(\beta)$ with $\rho(\beta)$ being its
normalised probability distribution. The space   
$\left\lbrace \left| \beta\right\rangle \right\rbrace$ has a size of $2^N$ where $N$ is the number of sites and hence we cannot calculate exactly the expectation value
using Eq.~\eqref{MCeq}. Therefore we use a Monte Carlo algorithm for this purpose\cite{Gros}.
We start the walk with a state $\left|\beta\right\rangle$, initialized through randomly filling up and down spins in the position space,  and derive a
new state $\left|\beta'\right\rangle$ at
each step by flipping a pair of randomly chosen spins. At each step, the new overlap amplitude
$\left|\left\langle \beta'|\psi_{SF} \right\rangle \right|^2$ given by a Slater determinant is calculated. 
The ratio between the new and old overlap amplitudes is used as the acceptance ratio.
After every $N$ steps, a measurement of the function $f(\beta)$, as defined in Eq.~\eqref{MCeq},
is performed by calculating the matrix elements $\left\langle \beta \right| O \left| \gamma \right\rangle$
and the overlap $\left\langle \gamma \right| \left. \psi_{SF} \right\rangle$.
In principle the sum runs over all states $\left|\gamma \right\rangle$ in the single-occupancy basis, but since we are only interested in averaging local operators $O$  (such as energy or static spin structure factor),
for a given $\left|\beta \right\rangle$ there are only a few relevant $\left|\gamma\right\rangle$ states
with nonzero  $\left\langle \beta \right| O \left| \gamma \right\rangle$.
For calculating the energy, the operator $O$ is the physical Hamiltonian Eq. \eqref{H-physical}.

\section{Results}
\subsection{Variational Energy}
\begin{figure}[H]
    \centering
    \includegraphics[scale=0.7]{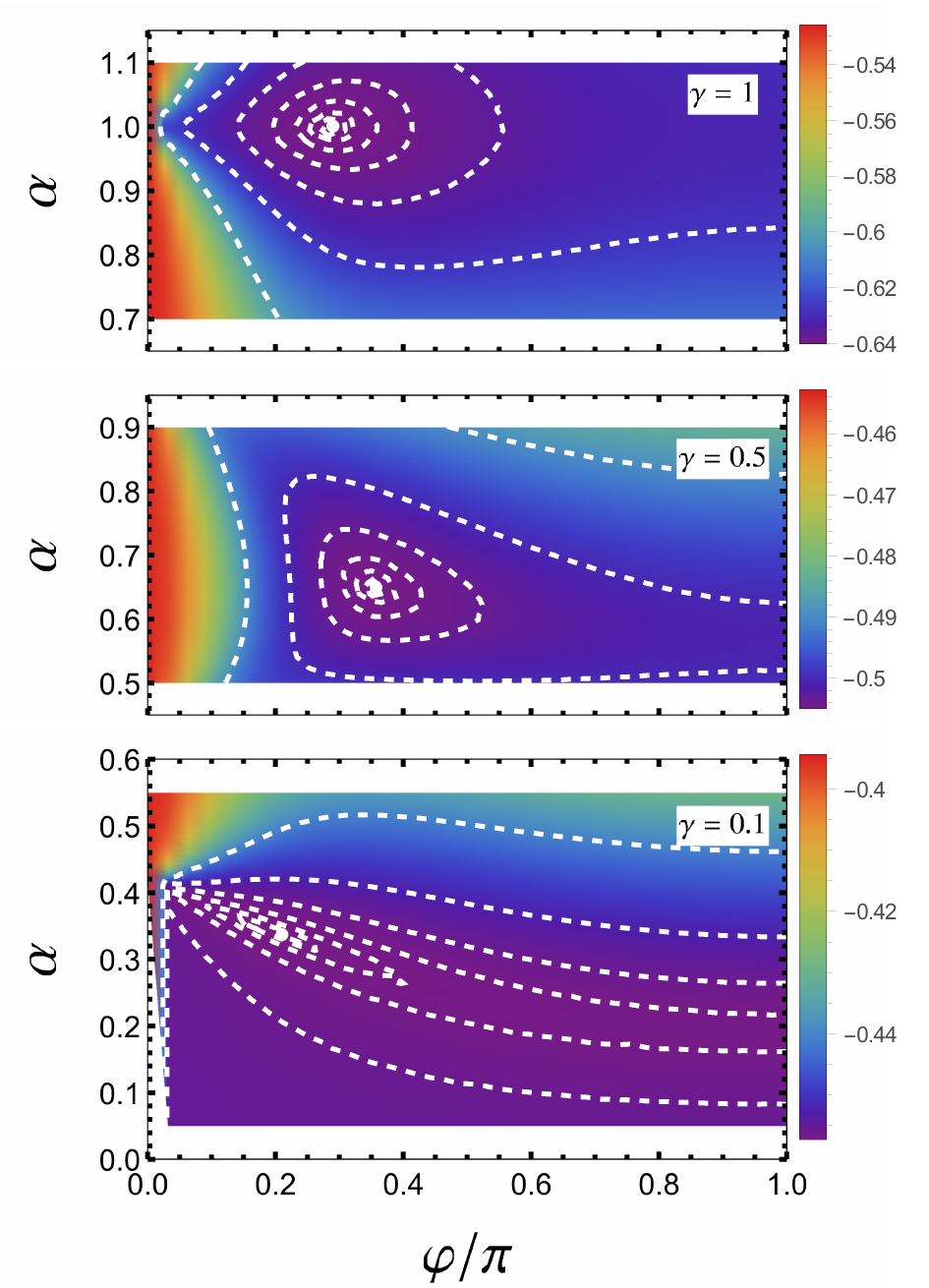}
    \caption{Ground state energy maps of systems with size $L=8$ for various coupling ratios. The white contours (dot) indicate the points corresponding to same energy (minima)}
    \label{fig:minima}
\end{figure}

\begin{figure}[H]
\centering
    \includegraphics[width=\linewidth]{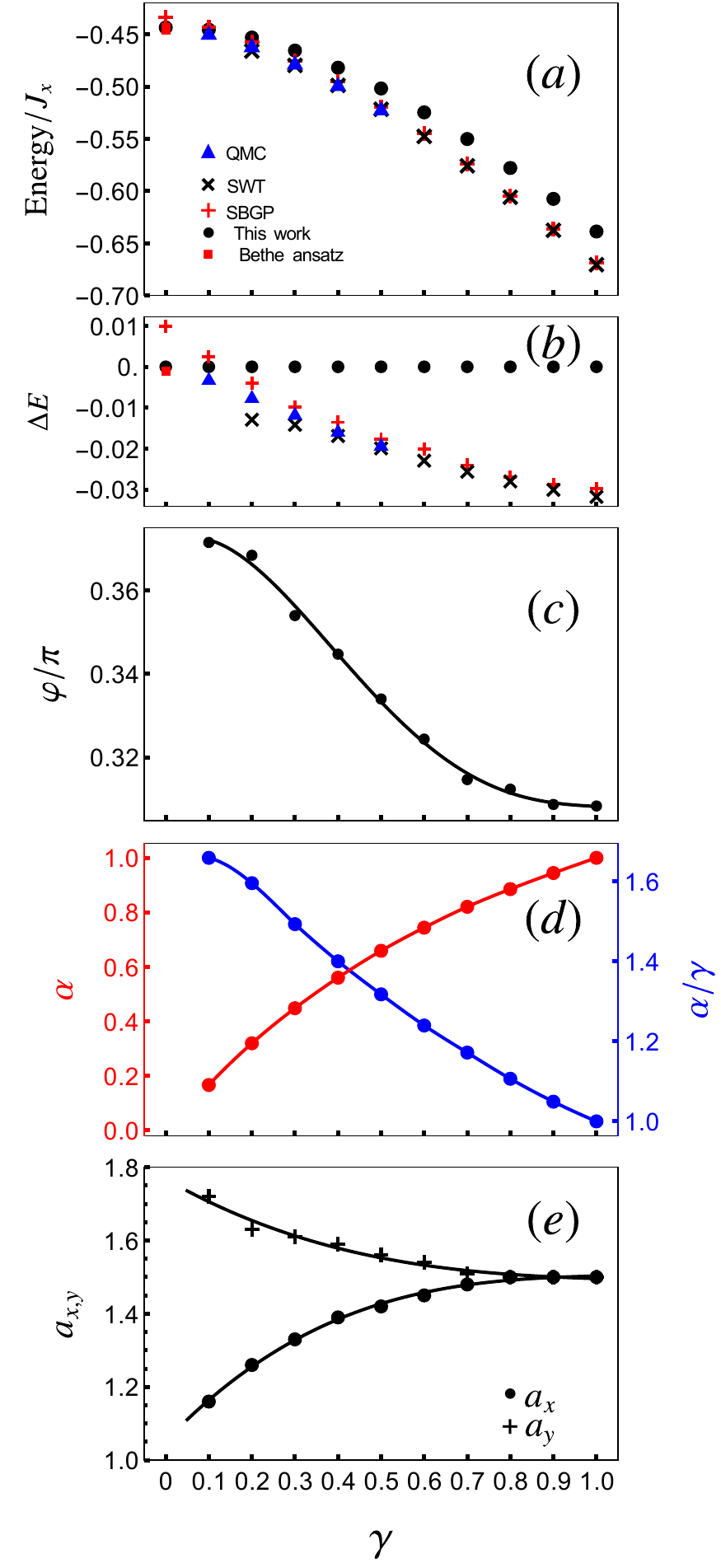}
\caption{Properties of the variational wave function as function of $\gamma$.
 (a) Energies and their (b) differences, compared to quantum monte carlo methods (QMC), spin wave theory (SWT), Gutzwiller projected Schwinger boson states (SBGP) and Bethe ansatz (at $\gamma =0$). Optimized variational parameters $\varphi$ (c) and $\alpha$ (d). (e) Exponents of the algebraically decaying staggered correlation functions. The lines are guides to eye.}
    \label{fig:varpar}
\end{figure}

The first step of the calculation is optimizing the values of  variational parameters $\alpha$ and $\varphi$ by finding the minimum energy of the variational state. FIG.~\ref{fig:minima} shows the energy maps in $(\varphi,\alpha)$ parameter space at $\gamma =\{0.1,0.2,0.5,1\}$ for system size $L=8$. At large $\gamma$, the minima are well defined, but closer to $\gamma \sim 0$ the minima become shallow in $\varphi$. This comes as no surprise, since when approaching the 1D case, the notion of flux around a loop is ill-defined. The ground state energies and the corresponding  optimum parameters, for system sizes $L\times L$ with $L=\{8,12,16,24\}$, were extracted by fitting the low energy part of the maps with a quadratic function in $\varphi$ and $\alpha$. The optimum parameters and energy at the system size $L=24$ are shown in FIG.~\ref{fig:varpar}. The ground state energy is compared with the variational work by Miyazaki et al.\cite{Miyazaki} using Gutzwiller projected schwinger boson states (SBGP), Quantum Monte Carlo\cite{Sandvik,sandvick2}(QMC), and spin wave theory (SWT)\cite{Shaik} including the linear part and next order corrections. Starting from the symmetric case $\gamma =1$, we observe that the staggered flux state has higher energy compared to all three methods. The energy difference decreases as coupling ratio is decreased, and at $\gamma \leq 0.1$ we observe SF state outperforming the SBGP result. With decreasing $\gamma$, the flux parameter $\varphi$ increases slightly and the amplitude ratio $\alpha$ decreases. Interestingly, $\alpha$ decreases slower than $\gamma $, such that $\alpha/\gamma$ increases with decreasing $\gamma$, as shown in FIG.~\ref{fig:varpar}(d).

\subsection{ Finite size analysis}
Since our calculations are performed on finite lattices, a finite size analysis is needed to establish the convergence of our parameters and validate our conclusions for infinite size limit. It must be noted here that our ground state wave function explicitly depends on the phase of the function in Eq.~\eqref{Delta} which is ill-defined at the nodal point. To avoid this point, for system sizes $L=4n$, Dalla Piazza et.\ al.\cite{DallaPiazzath,DallaPiazza} worked with anti-periodic boundary conditions in x and y directions, termed here as \textit{abc}-\textit{abc} (used in FiG.~\ref{fig:varpar}). We take this method one step further, by including another possibility, periodic in x and anti-periodic in y (\textit{pbc}-\textit{abc}).

The advantage of using different boundary conditions is that it provide us with an efficient method to identify features explicitly related to finite system size as opposed to ones that can be extended to $L\rightarrow\infty$.  The ground state energy calculations are also performed on system sizes of type $L=4n+2$ with boundary conditions \textit{pbc}-\textit{pbc},\textit{abc}-\textit{pbc} and the optimum parameters derived through a linear extrapolation of the $L=4n$ parameters.

\begin{figure}[H]
\captionsetup[subfigure]{labelformat=empty}
\def\tabularxcolumn#1{m{#1}}
\includegraphics[scale=0.75]{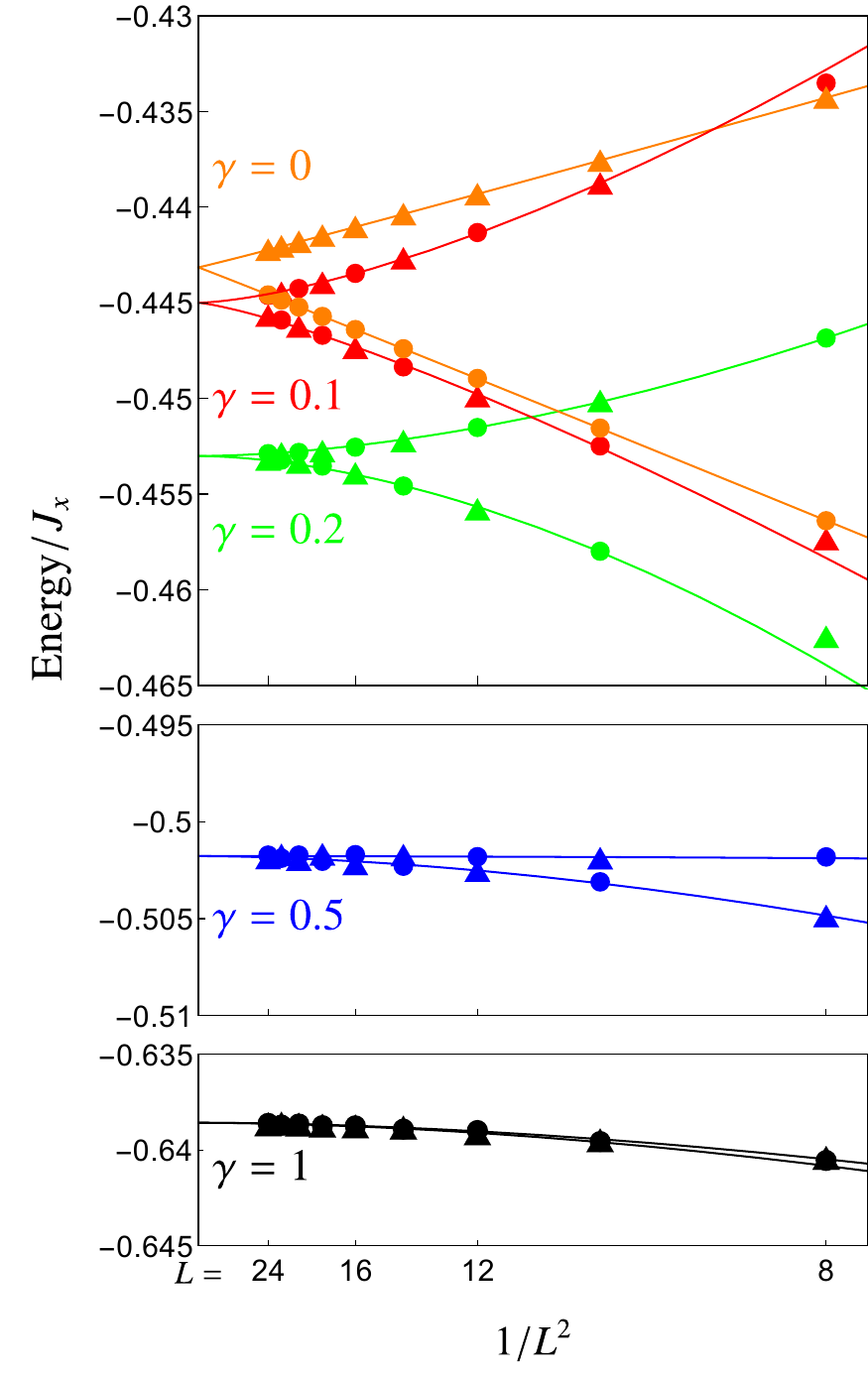}
\caption{ Finite size scaling of ground state energies along with the algebraic fitting for selected $\gamma$. The plot markers are assigned with respect to boundary condition in x-direction representing \textit{abc}(\textit{pbc}) with triangles(circles).  The $\gamma=0$ plots and fits correspond to exact solution using Bethe ansatz.}
\label{fig:fs}
\end{figure}

FIG.~\ref{fig:fs} shows the energies corresponding to the mentioned possibilities. To avoid confusion from here on we refer to the boundary condition only in x-direction, and it is understood that corresponding boundary conditions in y-direction are \textit{abc} for $L=4n$ and \textit{pbc} for $L=4n+2$. From FIG.~\ref{fig:fs}, we can see that the sensitivity to boundary conditions decreases with increasing coupling. A comparison with the energies from Bethe ansatz shows an important observation. For systems $L=4n$, we observe that our wavefunctions with \textit{abc}, are closer to the Bethe ansatz with \textit{pbc}, and vice -versa. This situation is reversed for $L=4n+2$. The observed equivalence between \textit{pbc} (\textit{abc}) for $L=4n+2$ and \textit{abc} (\textit{pbc}) for $L=4n$ can be easily understood by inspecting the corresponding $k$ space where $k_x=\pi/2$ is avoided (included). The equivalence between \textit{abc} for $L=4n$ and \textit{pbc} for Bethe ansatz  is due to the fermionic nature of our wavefunction which upon the imposition of translational symmetry incur a sign difference that depends on whether there is an even or odd number of down spins.

 \subsection{Instantaneous spin correlation}
\begin{figure}[H]
    \centering
    \includegraphics[scale=0.7]{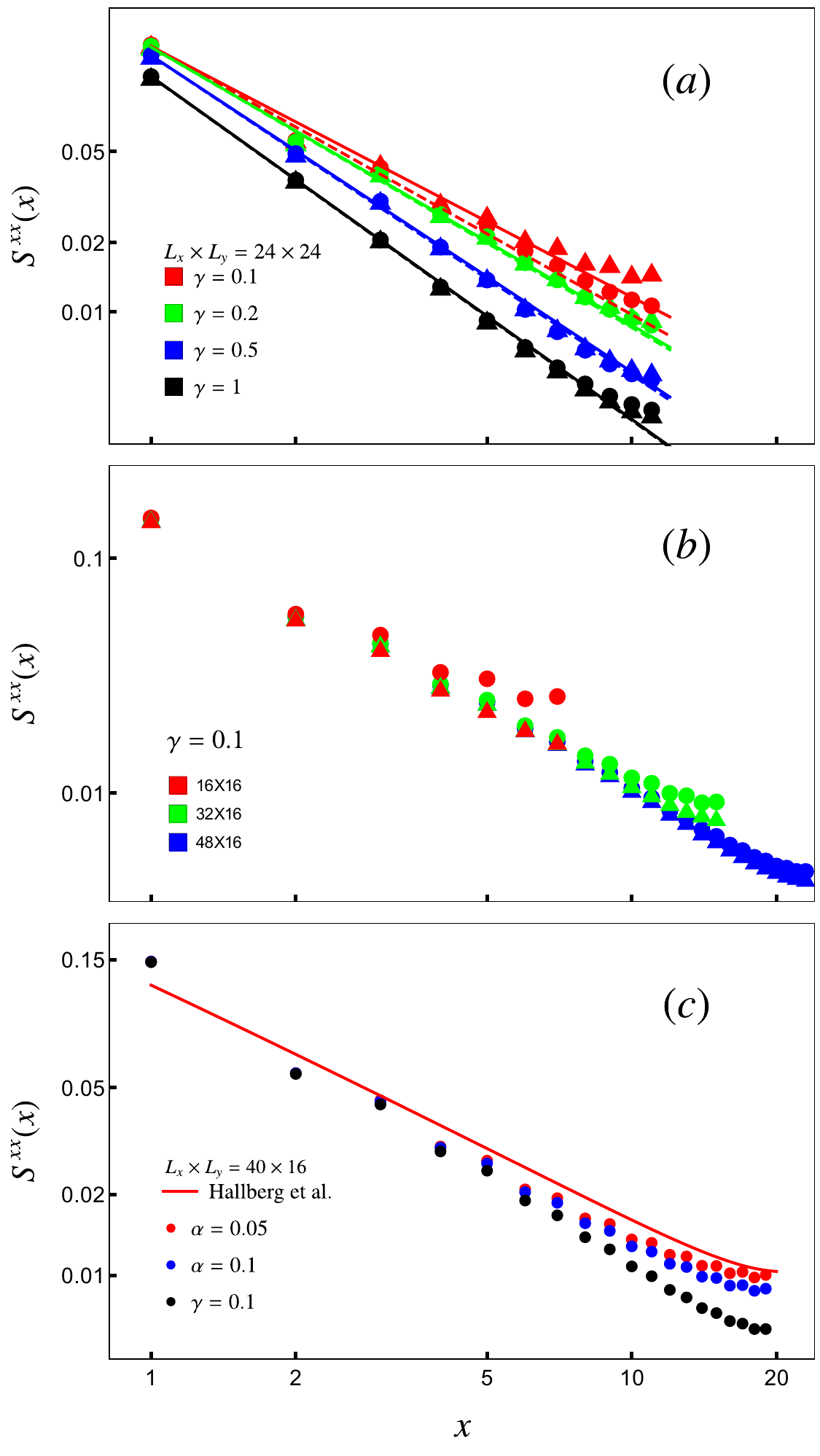}
\caption{(a) $S^{xx}(x)$ for $\gamma=0.1,0.2,0.5,1$  and the corresponding fits to algebraic decay $S^{xx}(x)\propto x^{-a_x}$ at $L=24$. (b) $S^{xx}(x)$ at $\gamma=0.1$ for different system sizes. (c) $S^{xx}(x)$ for  $\alpha=0.05,\alpha=0.1,\gamma=0.1$ for system size $40\times16$ compared to expression from Hallberg et al. \cite{Hallberg}.The plot markers in (a,b) are assigned with respect to boundary condition in x-direction representing \textit{abc} (\textit{pbc}) with triangles (circles).}
    \label{fig:ssfx_all}
\end{figure}

Next, we calculate the is the instantaneous staggered spin-spin correlation function $S^{\alpha\alpha}(r)=1/N\sum_i e^{i Q.r}\left\langle S^\alpha_{i+r}S^\alpha_i\right\rangle$ with $Q=(\pi,\pi)$. In the absence of symmetry breaking long range order, $S^{xx}(r)$, $S^{yy}(r)$ and $S^{zz}(r)$ are equivalent. Numerically, we observe that the $xx$ component converges faster than the $zz$ component (see supplementary). Starting with the x direction, as can be seen in FIG.~\ref{fig:ssfx_all}(a), $S^{xx}(x)$ decays as a power-law. At large $\gamma$, the correlation functions are insensitive to the boundary conditions, while at small $\gamma$ (e.g.\ $\gamma$=0.1), deviations appear noticeable at large $r$. 
This is solved by increasing system size. As shown for $S^{xx}(x)$ at $\gamma=0.1$ in FIG.~\ref{fig:ssfx_all}(b), when increasing $L_x$ with $L_y=16$, the correlation functions converge. 
Interestingly, as can be seen from the slopes in FIG.~\ref{fig:ssfx_all}(a), the exponent $a_x$ in $S^{xx}(x) \propto x^{-a_x}$  appears to vary as a function of $\gamma$, as summarized in FIG.~\ref{fig:varpar}(e).

Various estimates exist for $S^{xx}(x)$ in the pure 1D case $(\gamma=0)$. The exact values for nearest neighbor\cite{Hulthen} $(0.14771)$ and the next nearest neighbor\cite{Takahashi} $(0.06068)$ are known. From field theory\cite{Affleck} the exact expression at $r\rightarrow \infty$ is $\sqrt{\ln r}/((2\pi)^{3/2}r)$. Results for finite system sizes have been calculated through density matrix renormalization group (DMRG) methods\cite{Hallberg}. To compare with these results, we calculate the correlation function at $\gamma\sim 0$ for larger system sizes $L_x\times L_y=40\times16$. Optimizing the variational parameters for $\gamma < 0.1$, is challenging since the energy minima are extremely flat in $(\alpha,\varphi)$ space.  We assume that $\alpha/\gamma$ has a finite value as we approch $\gamma=0$, and carry out the calculations for small $\alpha$ values. In FIG.~\ref{fig:ssfx_all}(c), we present the results for $\alpha=0.05,0.1$ and also for the optimized wave-function at $\gamma=0.1$. The nearest and next nearest neighbor values at $\alpha=0.05$ are $0.1475$ and $0.05662$, close to the exact values. A comparison with the DMRG result from the work of Hallberg et al\cite{Hallberg}, shows that our correlation functions at $\gamma \sim 0$ progressively get closer to the 1D estimate as we reduce the coupling ratio $\gamma$.

\begin{figure}[H]
    \centering
    \includegraphics[scale=0.7]{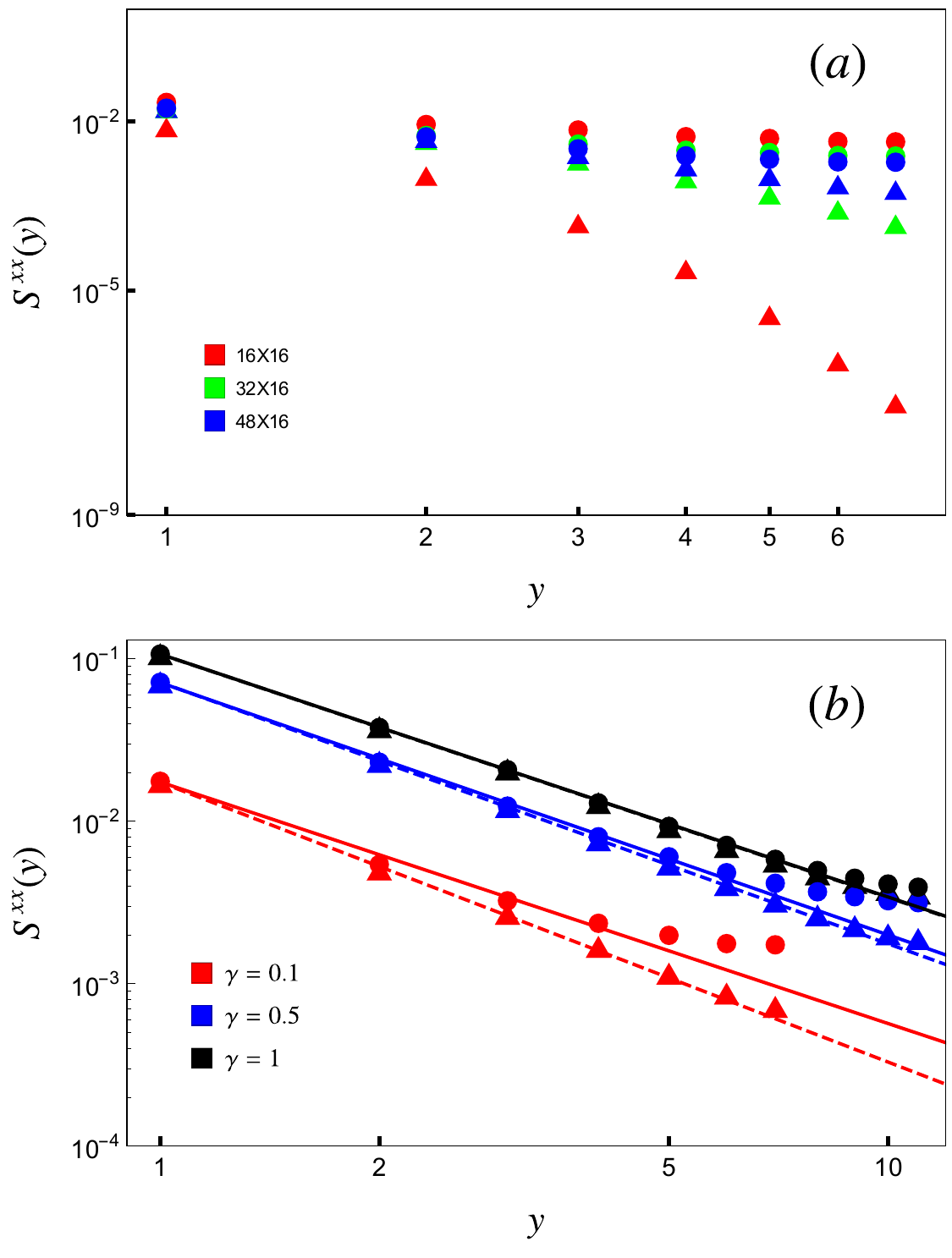}
    \caption{(a) Sensitivity of correlations in y-direction to the boundary conditions at $\gamma=0.1$ for different system sizes. (b) Correlations in y-direction for couplings $\gamma=0.1,0.5,1$ and the corresponding fits to algebraic decay $S^{xx}(y)\propto y^{-a_y}$ at $L=24$. The plot markers are assigned with respect to boundary condition in x-direction representing \textit{abc} (\textit{pbc}) with triangles (circles). }\label{fig:ssfy_all}
\end{figure}

Along the $y$-direction, the correlation function becomes very sensitive to boundary conditions for small $\gamma$. Interestingly, the remedy is to increase system size along the strong coupling $x$-direction as shown in FIG.~\ref{fig:ssfy_all}(a). Similar to the $x$-direciton, the correlation functions remain algebraic, however with an exponent $a_y$ that increases with decreasing $\gamma$. The fact that $a_y$ increases and $a_x$ decreases with decreasing $\gamma$ would imply that as coupling between chains weaken, correlations decay faster across chains, but decay slower along the chain than in the 2D square lattice case.
We note that the exponents extracted here are fits of up to 12 lattice spacings. It is possible that the asymptotic exponents in the large distance limit would behave differently, for instance converge to a universal value.

 \section{Discussion}

Our work focuses on extending the staggered flux variational wave-function approach to rectangular lattices, and the ground state properties of this wave function have been presented in this article. In the main part of the paper, we have compared our results with other methods, and in this section we discuss the key conclusions on this comparison. Starting with the ground state energy, for large $\gamma$, our estimates are higher than the QMC\cite{Sandvik}, SBGP\cite{Miyazaki} and SWT results\cite{Shaik}. This is not surprising, and has been already noted by Dalla Piazza et al\cite{DallaPiazza,DallaPiazzath}. A disordered staggered flux state for the square lattice has higher energy ($-0.638J$)  compared to an ordered staggered flux state ($-0.664J$) and the current best estimate by the Green's function Monte Carlo method ($-0.669J$)\cite{Trivedi,Runge,Calandra}. However, they also note that, although the ordered staggered flux state performs better energetically, it does not reproduce the quantum anomaly, exhibits a gapped-spectrum, and has exponentially decaying spin-spin correlations contrary to the expected power-law decay.
 
On the other end of the coupling ratio is the quasi-1D case $\gamma \rightarrow 0$ where the staggered flux wave function energetically performs better than the SBGP state but has slightly higher energy than the QMC result. To estimate the energy in the pure 1D case, Miyazaki et al\cite{Miyazaki} set $J_y=0$ and treat $\gamma$ as a variational parameter, and the optimum result yields $E(\gamma=0)=-0.4337$ at $L=20$. Using a similar logic, at $L=20$ we set $\gamma=0$ and calculate the energy of a state with very small $\alpha=0.05$. This  yields value of $E(\gamma=0)=-0.4442$, which is very close to the exact value from Bethe ansatz $E(\gamma=0,L=20)=-0.4445$.
It is believed that from the limit of coupling spin-chains, long range order sets in already at infinitesimal inter-chain coupling. On the other hand, the ordered moment calculations through spin-wave theory (with first correction) reaches zero for $\gamma=0.138$, below which spin wave theory breaks down. Though similar observation was made through a mean field treatment by Miyazaki et al., where the ordered moment goes to  zero at a value $\gamma =0.1356$, the analysis of SBGP state at $\gamma =0$ seems to indicate that long-range order exists all the way down to $\gamma=0$. This suggests that the loss of order at $\gamma \sim 0.138$ is just an artefact of the mean field methodologies. Interestingly in our work at $\gamma \leq 0.1$, the ground state energy of the staggered flux state is lower than the SBGP result. This indicates that, although the staggered flux result at $\gamma=0.1$ falls short of outperforming the QMC result, within the framework of variational wavefunctions, the staggered flux fermionic wavefunction outperforms the bosonic SBGP wavefunction. In conclusion, while the lack of N\'eel order compromises the staggered flux state at high $\gamma$, it appears to capture the essence of the system at low $\gamma$. We therefore believe the presented staggered flux state provides an interesting starting point for exploring the cross-over from quantum disordered chains to the N\'eel ordered 2D square lattices. 

\section*{Acknowledgements}

We would like to thank Bowen Zhao and Anders W. Sandvik for the sharing the QMC data included in FIG.~\ref{fig:varpar}. 

\bibliography{bibliography}
\bibliographystyle{apsrev4-1}

\clearpage
\onecolumngrid
\section*{Supplementary Information}

\subsection*{Optimization}
\begin{figure}[H]
  \centering
  \begin{minipage}[b]{0.3\textwidth}{
    \includegraphics[width=\textwidth]{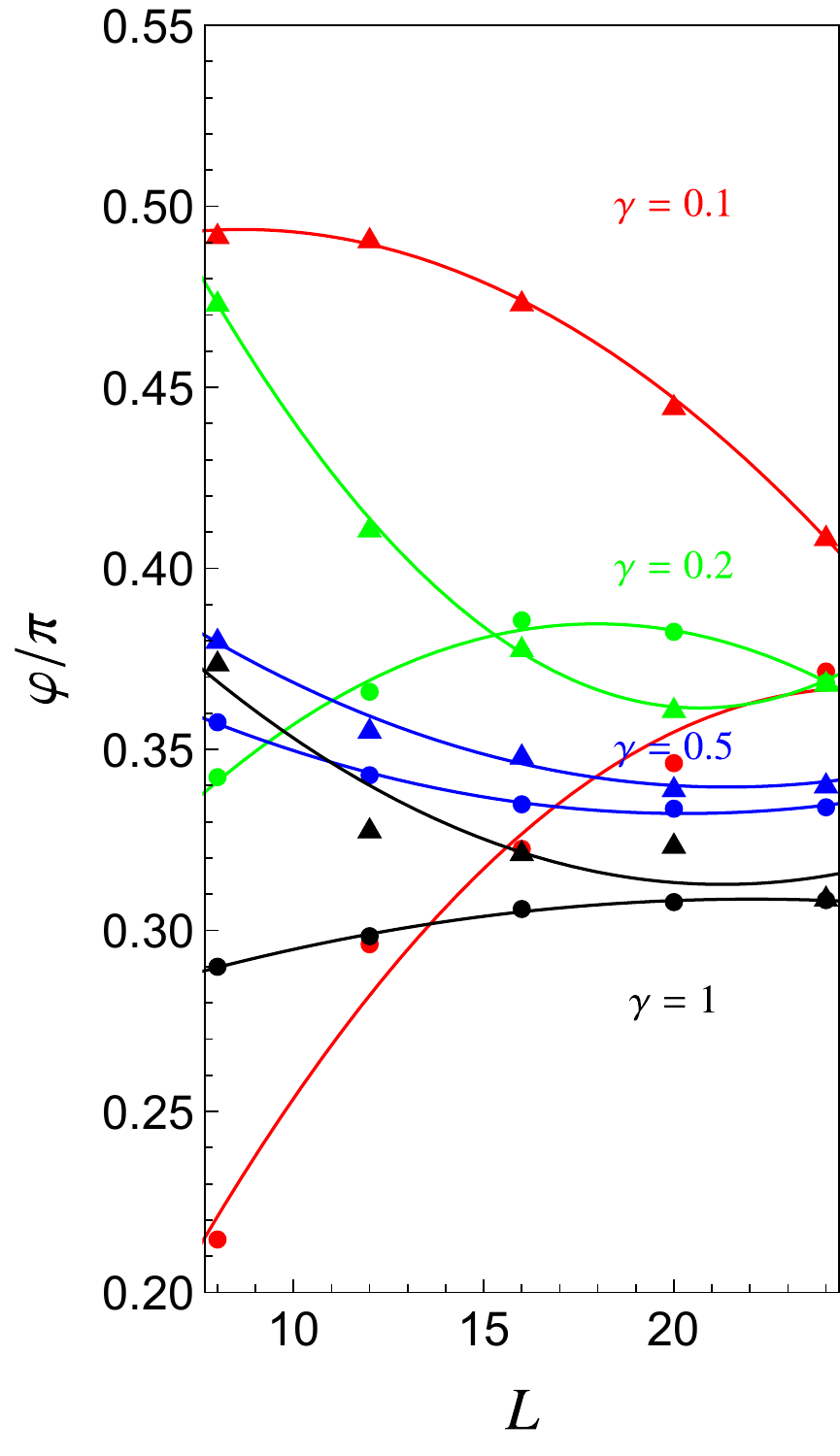}}
  \end{minipage}\hspace{1cm}
  \begin{minipage}[b]{0.3\textwidth}{
    \includegraphics[width=\textwidth]{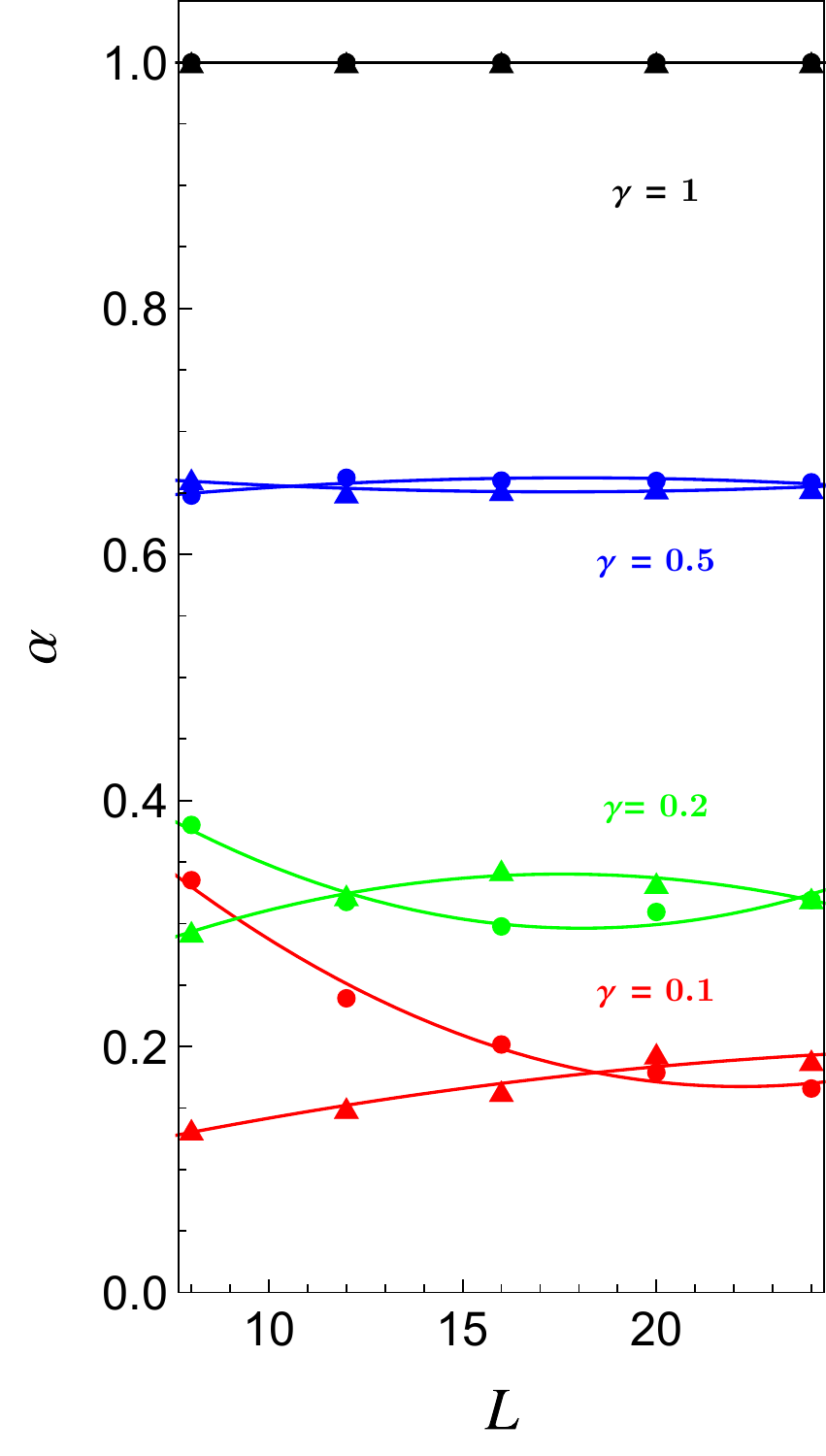}}
  \end{minipage}
  \caption{Parameters $\alpha$ and $\varphi$ corresponding to FIG.~\ref{fig:fs}. The plot markers are assigned with respect to boundary condition in x-direction representing \textit{abc} (\textit{pbc}) with triangles (circles). The lines are guides to eye.}\label{fig:fs_params}
\end{figure}

To study the finite size effects, the energy maps were calculated for system sizes $L=8,12,16,20,24$ for both boundary conditions. To minimize the computational cost, energy maps over large parameter space (as shown in FIG.~\ref{fig:minima}) were calculated only at $L=8$, and for subsequent system sizes the energy calculations were performed on smaller regions in the parameter space. These regions are selected through the estimates from previous system sizes and are verified to contain the energy minimum. As mentioned in the article, the optimum parameters are derived by fitting the lower part of the energy minimum with a second order polynomial in  $\alpha$ and $\varphi$. The optimum parameters are plotted in FIG.~\ref{fig:fs_params}.

As can be seen from above, we do not observe a monotonic behaviour between our estimates of ($\alpha$,$\varphi$) vs $L$. This is partly due to the fact that our parameter estimates are subject to errors caused due to the fitting of energy minima with second order polynomial in $\alpha$ and $\varphi$, whereas our energy maps are not exactly parabolic. Nevertheless, we see a qualitative convergence of our parameters and the corresponding energy estimates are observed to have a very small error($<10^{-4}$). 

\subsection*{Parameters}

\begin{table}[H]
\begin{center}
  \resizebox{0.85\columnwidth}{!}{ \begin{tabular}{|*{11}{c|}}
\hline
     $\gamma$&0.1&0.2&0.3&0.4&0.5&0.6&0.7&0.8&0.9&1\\
     \hline
     $\varphi(\text{in}\ \pi)$&0.371&0.368&0.354&0.345&0.334&0.325&0.315&0.312&0.309&0.308\\
     \hline
     $\alpha$&0.166&0.319&0.448&0.560&0.659&0.74&0.820&0.885&0.944&1\\
     \hline
     $E(L=24)$& -0.446 &-0.453&-0.466&-0.482&-0.502&-0.525&-0.550&-0.578&-0.607&-0.639\\
 \hline
     $E(L=\infty)$& -0.4450 &-0.4530&-0.4654&-0.4819&-0.5018&-0.5246&-0.5501&-0.5778&-0.6074&-0.6385\\
 \hline
\end{tabular}}
\end{center}
\caption{Optimized variational parameters and corresponding Energy at $L=24$. and estimates at $L=\infty$ for different values of $\gamma$}\label{param_tab}
\end{table}
In TABLE I, we present the numerical values of the energy minima, and the corresponding optimum parameters used in FIG.~\ref{fig:varpar}(a,c,d) of the article.  These parameters correspond to the $L=24$ system size with \textit{abc} boundary conditions. 

\subsection*{Correlation funcitons}

\begin{figure}[H]
  \centering
  \begin{minipage}[b]{0.4\textwidth}{
    \includegraphics[width=\textwidth]{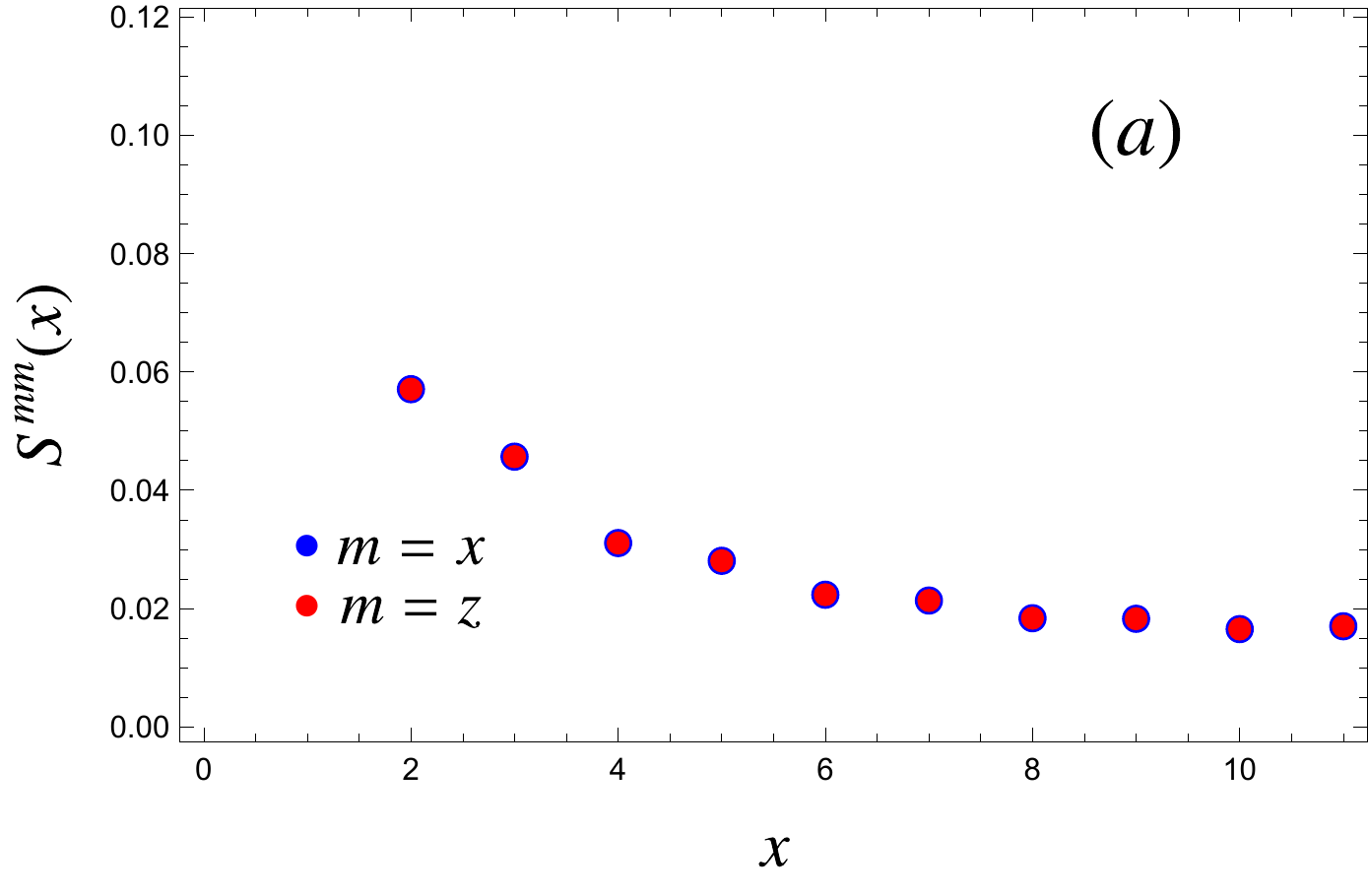}}
  \end{minipage}\hspace{1cm}
  \begin{minipage}[b]{0.4\textwidth}{
    \includegraphics[width=\textwidth]{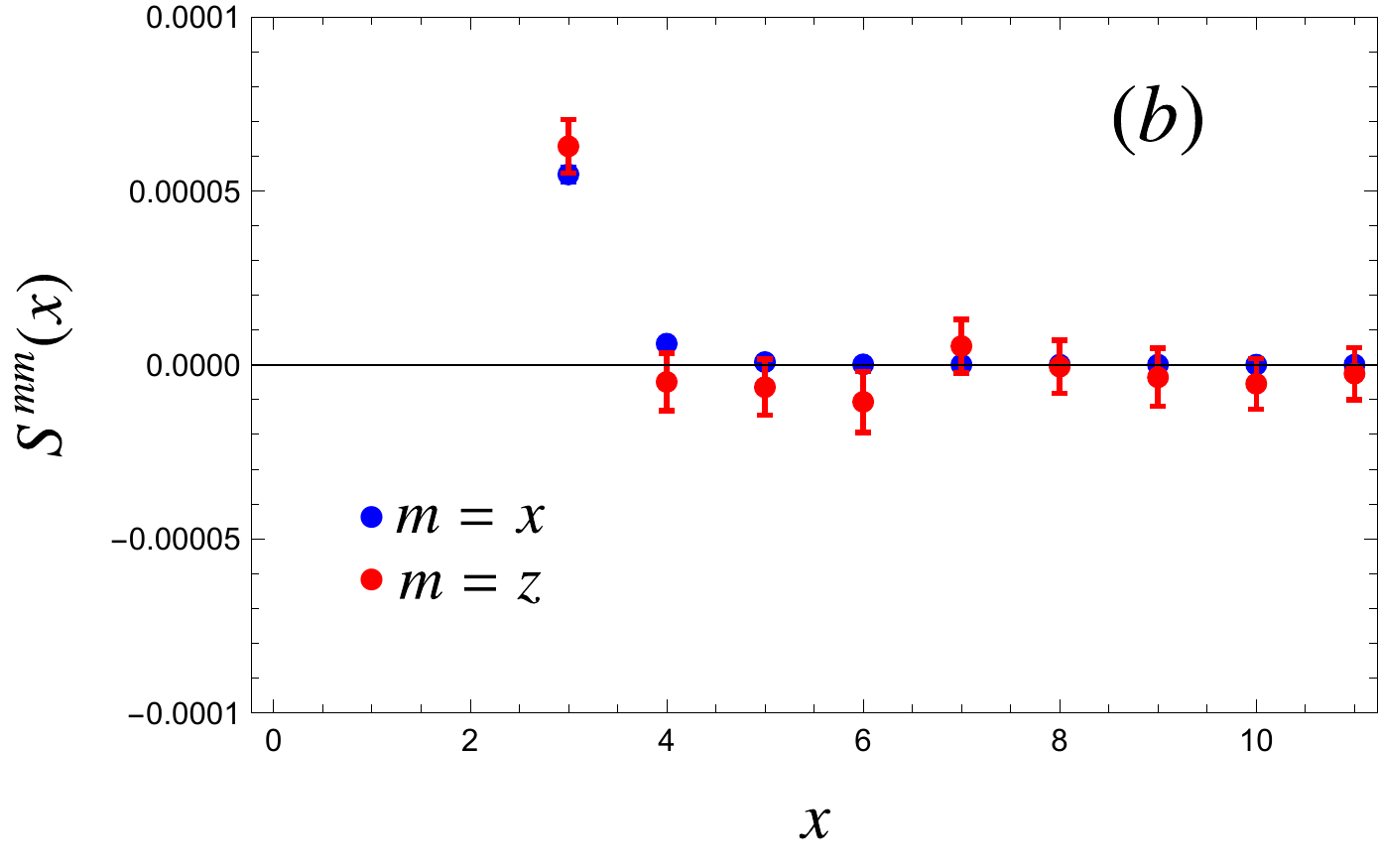}}
  \end{minipage}
\caption{Instantaneous spin spin correlation function along with the standard error along x(a) and y(b) directions for simulation at $\alpha=0.1$. }\label{fig:errssf}
\end{figure}

For a disordered state like the staggered flux state, the $xx$ and $zz$ components of the instantaneous staggered spin correlation function are expected to have the same values. However, the fact that we estimate these values through a statistical numerical process i.e a Monte-carlo process in a space defined via spins pointing up and down in z-direction, we expect a small difference in our estimated values. Luckily, this difference is minimal (practically invisible in FIG.~\ref{fig:errssf}(a)) in most of the cases and appears only at points where the correlation function has very small values as in FIG.~\ref{fig:errssf}(b).

We also observe that the transverse components $xx$ converge faster than the $zz$ components as indicated by the standard error in FIG.~\ref{fig:errssf}(b). This is due to the fact that the calculation of transverse component via Eq.~(10), involves inclusion of more states than the longitudinal component. On the other hand, since it involves calculating off diagonal elements, the transverse component is more computationally demanding at each step compared to the longitudinal component, that involves only diagonal terms. For calculations corresponding to the instantaneous spin correlation functions at $L=24$ i.e FIG.~\ref{fig:ssfx_all}(a), FIG.~\ref{fig:ssfy_all}(b) same parameters as in FIG.~\ref{fig:fs_params} were used. For rectangular system sizes with $L_y=16$ and large $L_x$ i.e FIG.~\ref{fig:ssfx_all}(b) and FIG.~\ref{fig:ssfy_all}(a) the optimum parameters corresponding to $L \times L=24 \times 24$ were used. Approaching 1D limit while calculating correlation functions at $\alpha=0.05,0.1$, since the energy maps are extremely flat, we cannot estimate the optimum values for $\varphi$ with certainty. It was also observed that the correlation functions are weakly dependent on $\varphi$.  Hence, in FIG.~\ref{fig:ssfx_all}(c) we consider a constant value $\varphi=0.35\pi$ to qualitatively assess the correlation functions close to  1D limit. The exponents $a_x,a_y$ corresponding to fitting correlation function in x and y directions with  $f(x)=b_x/x^{a_x},f(y)=b_y/y^{a_y}$ that are shown in FIG.~\ref{fig:varpar}(e) are given by:
\begin{table}[H]
\begin{center}
    \resizebox{0.6\columnwidth}{!}{  \begin{tabular}{|*{11}{c|}}
\hline
     $\gamma$&0.1&0.2&0.3&0.4&0.5&0.6&0.7&0.8&0.9&1\\
     \hline
     $a_x$& 1.16 &1.26&1.33&1.39&1.42&1.45&1.48 &1.50 &1.50&1.50\\
 \hline
 $a_y$& 1.72 &1.63&1.61&1.59&1.56&1.54&1.51 &1.50 &1.50&1.50\\
 \hline
 
\end{tabular}}
\end{center}
\end{table}

For $\gamma \geq 0.6$, where the boundary effects are minimal, we fit the $L=24$ data to obtain the estimates for the exponents. For $\gamma < 0.6$ we use the data from rectangular system size $L_x\times L_y =56\times 16$, where we observe a good convergence of \textit{abc} and \textit{pbc} results, to estimate the exponents.

\end{document}